\documentclass[12pt,a4paper,oneside]{scrartcl}

\usepackage{graphicx}
\usepackage{amssymb}
\usepackage{amsmath}
\usepackage{subcaption}
\usepackage{authblk}

\usepackage[american]{babel} 
\usepackage[T1]{fontenc} 
\usepackage[utf8x]{inputenc}
\usepackage{txfonts}

\usepackage[pdftex,unicode]{hyperref}
\hypersetup{hidelinks}
\usepackage{url}
\usepackage[capitalize,noabbrev]{cleveref}

\title{\textbf{Periodicity Pitch Detection in Complex Harmonies on {EEG} Timeline Data}}

\author[1,2]{Maria Heinze}
\author[2]{Lars Hausfeld}
\author[2]{Rainer Goebel}
\author[1]{Frieder Stolzenburg}
\affil[1]{Department of Automation and Computer Sciences, Harz University of Applied Sciences, Friedrichstr.~57-59, 38855 Wernigerode, Germany}
\affil[2]{Department of Cognitive Neuroscience, Maastricht University, Minderbroedersberg 4-6, 6211 LK Maastricht, Netherlands}

\date{}							
\begin{document}
\maketitle

\begin{abstract}
An acoustic stimulus, e.g., a musical harmony, is transformed in a highly
non-linear way during the hearing process in ear and brain. We study this by
comparing the frequency spectrum of an input stimulus and its response spectrum
in the auditory processing stream using the frequency following response (FFR).

Using electroencephalography (EEG), we investigate whether the periodicity pitches of complex harmonies
(which are related to their missing fundamentals) are added in the auditory
brainstem by analyzing the FFR. While other experiments focus on
common musical harmonies like the major and the minor triad and dyads, we also
consider the suspended chord. The suspended chord causes tension foreign to the
common triads and therefore holds a special role among the triads.

While watching a muted nature documentary, the participants hear synthesized
classic piano triads and single tones with a duration of 300\,ms for the stimulus and
100\,ms interstimulus interval. We acquired EEG data of 64 electrodes with a
sampling rate of 5\,kHz to get a detailed enough resolution of the perception
process in the human brain.

Applying a fast Fourier transformation (FFT) on the EEG response, starting 50ms
after stimulus onset, the evaluation of the frequency spectra shows that the
periodicity pitch frequencies calculated beforehand $\pm$3\,Hz occur with some
accuracy. However, jitter turned out as a problem here. Note that the sought-for
periodicity pitch frequencies do not physically exist in the frequency spectra
of the stimuli.

\end{abstract}

\section{Introduction}

Hearing is one of the most important and enriching senses humans have. It
not only allows us to communicate, but it also offers us to immerse ourselves in
an abundance of emotions, concerning the musical aspect. Many people were
looking for a scientific, rational, and mathematical explanation of how
emotions caused by music work in our brains.
Numerous approaches tackle this question, studying the consonance/dissonance
of dyads or triads \cite{BK09,CF06,JKL12}. For instance, the major triad is often
associated with emotional terms like \emph{pleasant} or \emph{bright}, and, in
contrast to this, the minor triad with terms like \emph{sad} or \emph{dark}.
Empirical studies as well as EEG experiments reveal a clear preference order
on the perceived consonance/dissonance of common triads in Western music, e.g.,
major $\prec$ minor \cite{BK11,JKL12,SHP03}.


The periodicity of complex chords can be detected by the human brain
\cite{Lan97,Lan15}. We here concentrate on the detection of the periodicity
pitch of a chord which corresponds to the reciprocal of the period length of the
chord and can be derived from the physical waveform of the stimulus. The
periodicity pitch can be computed for every musical harmony and is related to
the missing fundamental frequency which usually is not present as tone component
in the stimulus. Periodicity pitch and tone pitch represent distinct dimensions
of harmony perception. The relative periodicity of a complex chord can be
determined as the approximated ratio of the period length of the chord relative
to the period length of its lowest tone component. The perceived consonance of a
harmony decreases as the relative periodicity increases~\cite{Sto15}.

\section{Aims}

The goal of this research is to develop a model how the human brain perceives
and processes musical sounds. For this, in our EEG experiments, six different
harmonies (triads) and four single tones are presented (cf.
\cref{tab:stim}). Concerning the triads we investigate whether the
periodicity pitches of complex harmonies \cite{Sto15} (related to their missing
fundamentals) are added in the auditory brainstem by analyzing the FFR
\cite{LS+09,LS+15}. Those experiments have been done in a similar way before by
Lee \emph{et~al.} \cite{LS+09,LS+15} with dyads and Bidelman and Krishnan
\cite{BK11} with triads and shall now be tested for reproducibility. Extending
our experiment with the not so often examined suspended chord, we also expect
some new insights regarding its dissolution by the major chord compared to other
subsequent chords.

\section{Related works}

Lee \emph{et~al.} \cite{LS+09,LS+15} demonstrate that acoustic periodicity is an
important factor for discriminating consonant and dissonant intervals. They
measure human auditory brainstem responses to four diotically presented musical
intervals with increasing degrees of dissonance and sought to explicate how the
subcortical auditory system transforms the neural representation of acoustic
periodicity for consonant versus dissonant intervals. They discover that the
phase-locking activity to the temporal envelope is more accurate (i.e. sharper)
in musicians than non-musicians. The intervals show the highest response in the
brainstem at about the periodicity pitch frequency (cf. \cite[Sect.~2.6]{Sto15}).

Lerud \emph{et~al.} \cite{LA+14} build on this work. They aim at explaining the
biophysical origin of central auditory nonlinearities. The nonlinear neural
transformation in the brain is studied by comparing the frequency spectrum
of the input stimulus and its response spectrum in the auditory brainstem. The
latter shows additional frequencies which are not present in the input spectrum,
in particular the periodicity pitch frequency. The authors introduce the concept
of mode-locking and find good correlation between the response spectra and their
model. Nevertheless, Stolzenburg \cite{Sto17b} suggests that there might be easier
explanations, namely the transformation of the input signal into pulse trains
(spikes) whose maximal amplitude is limited by a fixed uniform value.

Bidelman and Krishnan \cite{BK09} measure brainstem FFRs from nonmusicians in
response to the dichotic intervals. Neural pitch salience is computed for each
response using temporal autocorrelation and harmonic pitch sieve analyses.
Brainstem responses to consonant intervals are more robust and yield stronger
pitch salience than those to dissonant intervals. In \cite{BK11}, the same
authors measure the responses in the brain to four prototypical musical triads
(major, minor, diminished, augmented). Pitch salience computed from FFRs
correctly predict the ordering of triadic harmony stipulated by music theory.
The correlation between the ranking of neural pitch salience \cite[Figure~3]{BK09}
and periodicity is also significant \cite{Sto15}.

Ebeling \cite{Ebe07,Ebe08} presents a mathematical model to explain the
sensation of consonance and dissonance on the basis of neuronal coding and the
properties of a neuronal periodicity detection mechanism. This mathematical
model makes use of physiological data from a neuronal model of periodicity
analysis in the midbrain, whose operation can be described mathematically by
autocorrelation functions with regard to time windows. The mathematical model
makes it possible to define a measure for the degree of harmoniousness. This
procedure works well for dyads, but for triads and more complex chords the
correlation with empirical ratings is relatively low, which has already been
noticed in \cite[Sect.~2.5.3]{Ebe07}. 

Langner \cite{Lan92,Lan97,Lan15} assumes, since all frequency components of a
harmonic sound are multiples of its fundamental frequency, that the period of
the fundamental is also encoded in the cochlea in amplitude modulations
resulting from superposition of frequency components above the third harmonic.
As a consequence, the period of the fundamental is coded temporally in spike
intervals in the auditory nerve and analyzed by neurons in the auditory
brainstem cochlear nucleus. As already mentioned, referring to Langner's work,
\cite{Sto15} demonstrates that the perceived consonance of a harmony decreases
as the relative periodicity increases.

\section{Methods}

\subsection*{Subjects}
Seventeen healthy adult listeners (10 females, 7 males; mean age 31.4 years)
participated in this research. From all subjects an informed consent was
obtained. The Ethical Review Committee Psychology and Neuroscience of Maastricht
University approved this study.

\begin{table}
\centering
\begin{tabular}{lccc}
stimulus interval & musical pitches & frequency components (Hz) & periodicity pitch (Hz)
\\\hline \hline
G major 			& G3 		& 196 ~ 392		& 49 \\
root position 		& B3 		& 247 ~ 494		& \\
              			& D4 		& 294 ~ 587		& \\\hline
C major 			& G4 		& 392 ~ 784		& 131 \\
2nd inversion 		& C5 		& 523 ~ 1047		& \\
              			& E5 		& 659 ~ 1319		& \\\hline
G minor 			& G3 		& 196 ~ 392		& 20 \\
root position 		& B$$3 	& 233 ~ 466		& \\
              			& D4 		& 294 ~ 587		& \\\hline
G augmented 		& G3 		& 196 ~ 392		& 10 \\
root position 		& B3 		& 247 ~ 494		& \\
                       		& D$\sharp$4 	& 311 ~ 622		& \\\hline
G diminished 		& G3 		& 196 ~ 392		& 39 \\
root position 		& B$$3 	& 233 ~ 466		& \\
                      		& D$$4 	& 277 ~ 554		& \\\hline
G suspended 		& G3 		&196 ~ 392		& 33 \\
root position 		& C4 		& 262 ~ 523		& \\
                      		& D4 		& 294 ~ 587		& \\\hline
C 				& C3 		& 131\\\hline
C 				& C5 		& 523\\\hline
G 				& G2 		& 98\\\hline
G 				& G4 		& 392\\\hline \hline
\end{tabular}
\caption{Stimuli and the corresponding frequencies. The suspended chord used in
this experiment contains a perfect fourth.}
\label{tab:stim}
\end{table}

\begin{figure}
  \centering
  \begin{subfigure}[b]{0.32\linewidth}
    \includegraphics[width=\linewidth]{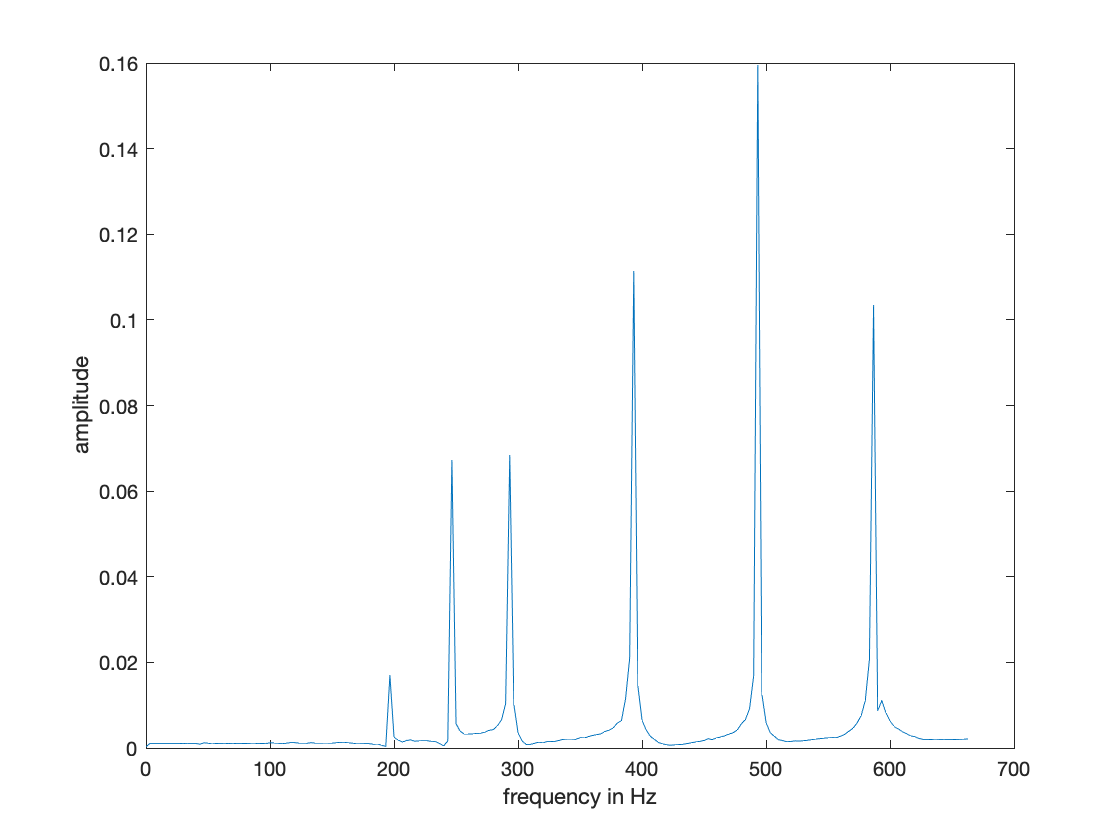}
    \caption{G3 major (root pos.)}
  \end{subfigure}
  \begin{subfigure}[b]{0.32\linewidth}
    \includegraphics[width=\linewidth]{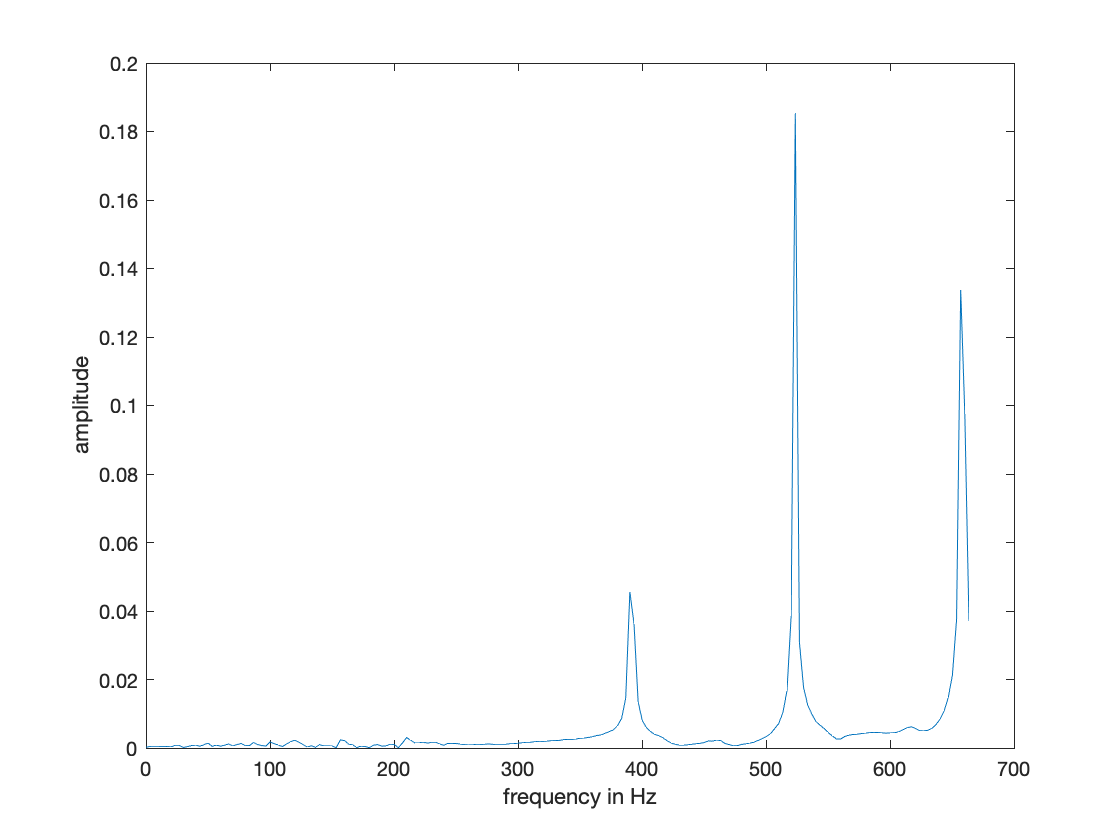}
    \caption{C4 major (2nd inv.)}
  \end{subfigure}
   \begin{subfigure}[b]{0.32\linewidth}
    \includegraphics[width=\linewidth]{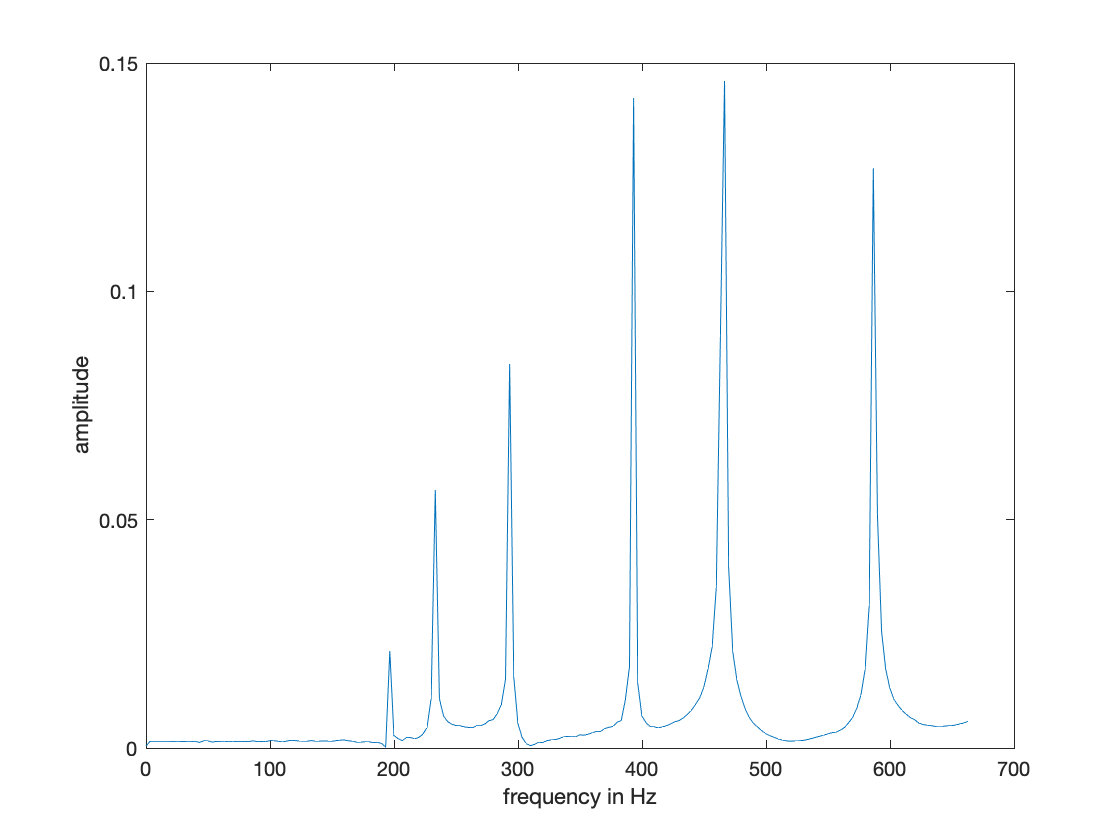}
    \caption{G3 minor (root pos.)}
  \end{subfigure}
  \begin{subfigure}[b]{0.32\linewidth}
    \includegraphics[width=\linewidth]{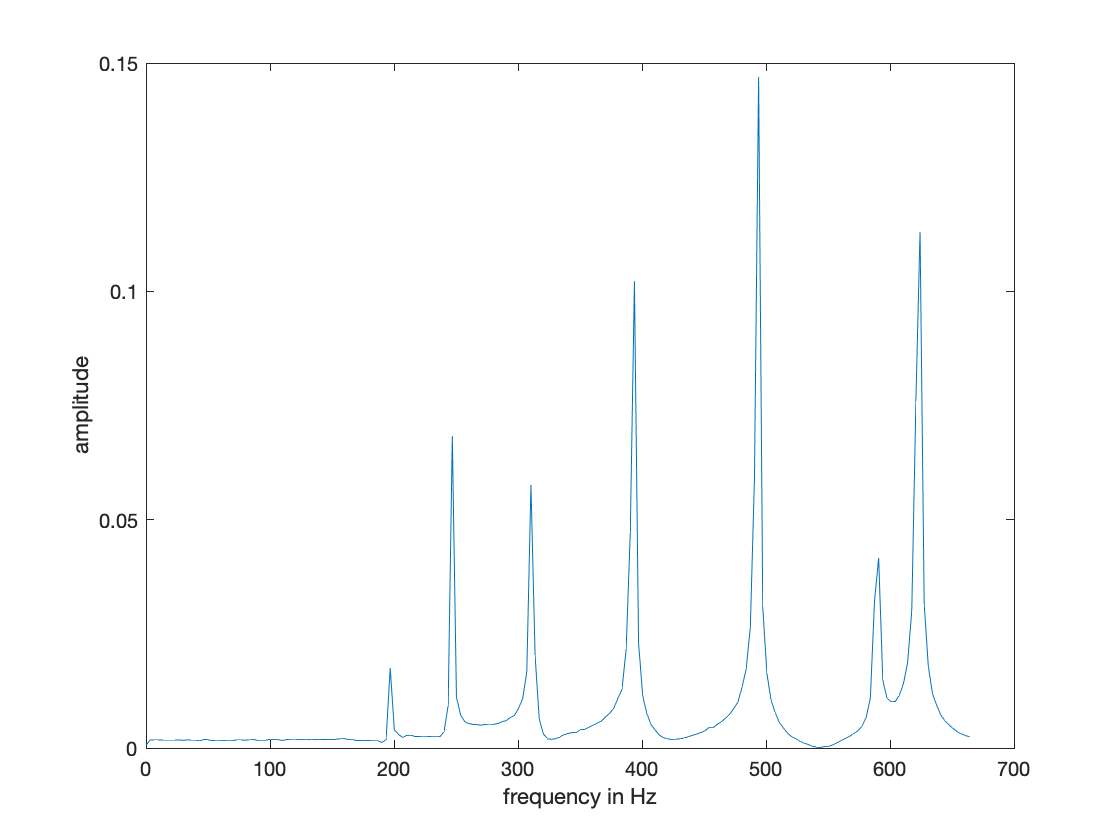}
    \caption{G3 augmented (root pos.)}
  \end{subfigure}
  \begin{subfigure}[b]{0.32\linewidth}
    \includegraphics[width=\linewidth]{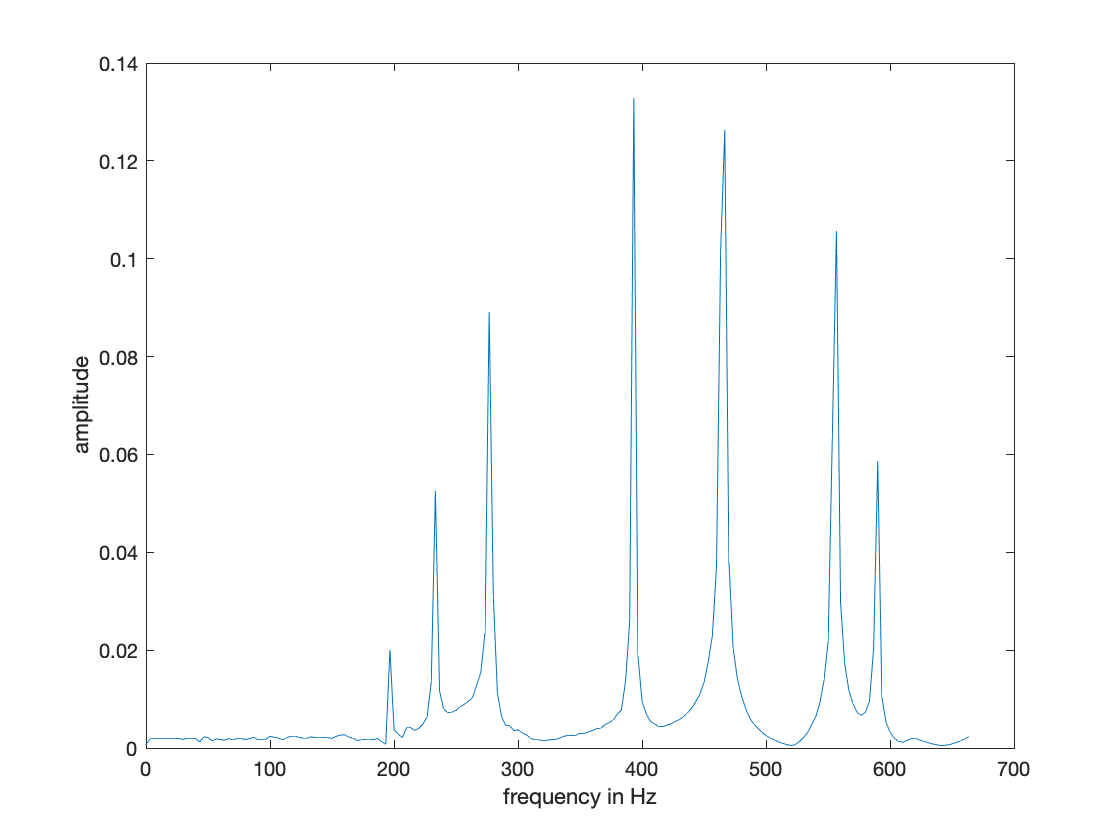}
    \caption{G3 diminished (root pos.)}
  \end{subfigure}
  \begin{subfigure}[b]{0.32\linewidth}
    \includegraphics[width=\linewidth]{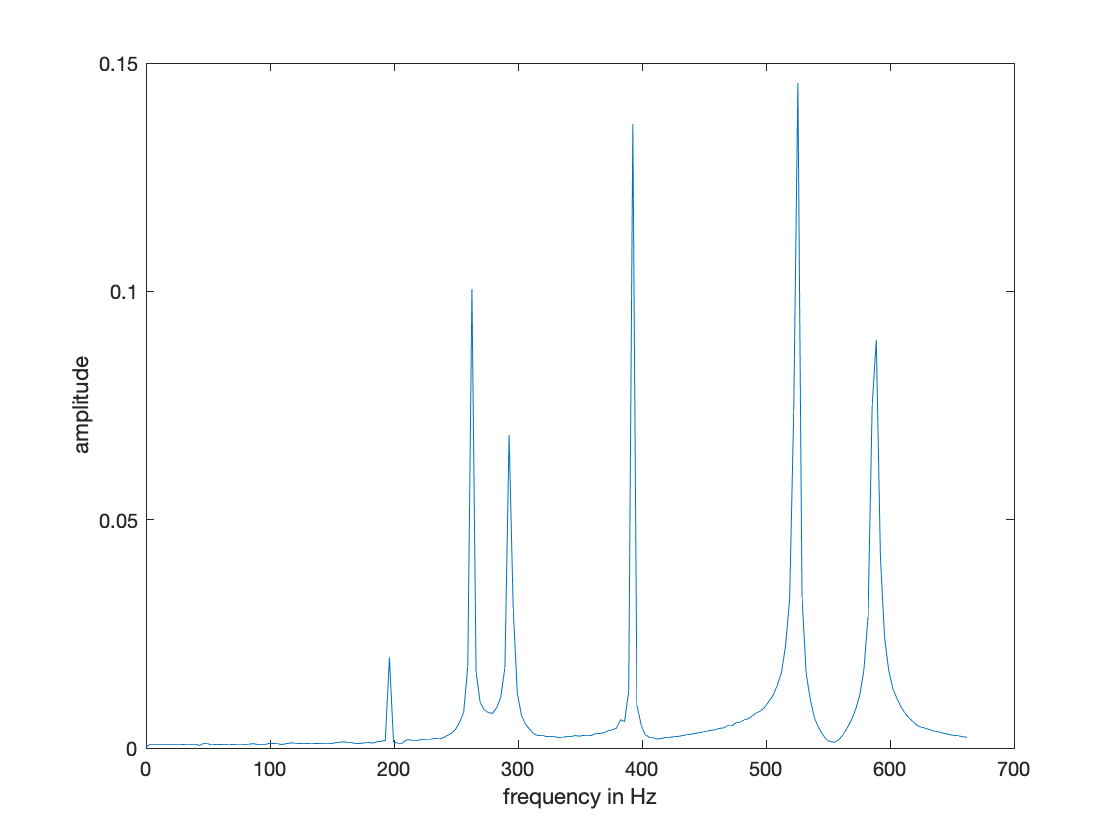}
    \caption{G3 suspended (root pos.)}
  \end{subfigure}
  \caption{Frequency spectra of the used triads.}
  \label{fig:stim}
\end{figure}

\subsection*{Stimuli}
A set of ten musical harmonies (6 triads and 4 single tones) were presented:
five of six triads were G chords in root position (major, minor, augmented,
diminished, and suspended), and one triad was a C major chord in its second
inversion (cf. \cref{tab:stim}). In addition, C3, C5, G2, and G4 were
presented as single tones. The timbre of the stimuli was a synthesized classic
piano sound with clear peaks in the corresponding fundamental frequencies (cf.
\cref{fig:stim}). The duration of each stimulus was 300\,ms. In all triads
the sought-for periodicity pitch frequencies did not physically exist in the
frequency spectra of the stimuli.

\subsection*{Procedure}
The EEG experiment procedure followed those in
\cite{LS+09,LS+15} and \cite{SK10}. The ten musical intervals
were presented with single polarity binaurally by loudspeakers in a soundproof
Faraday cage with an incoming intensity at around 67\,dB. The interstimulus
interval lasted 100\,ms. Responses were collected using Brain Products EEG
system with 64 electrodes. The contact impedance was $< 5$\,k$\Omega$ for all
considered electrodes. The musical intervals were each repeated 1,000 times in a
different order and their response recorded with a sampling rate of 5\,kHz.
While hearing, the participants watched a muted nature documentary.

\subsection*{Analysis}\label{analysis}
First all electrodes were re-referenced with the left mastoid electrode. The EEG responses were bandpass filtered in the range of 15–700\,Hz. If participants showed a greater activity than ±35\,µV, the corresponding trial was rejected. For the analysis two approaches from \cite{DKF05} were examined:
\begin{enumerate}
  \item the evoked method: All remaining trials and their baselines, the 50\,ms
	response before each trial, were averaged. An 
	FFT was performed on the averaged baselines as well as on the averaged
	trials (starting 50\,ms after stimulus onset). We examined the outcomes
	of four types of baseline correction \cite{LSJ05}, including the absolute, the
	relative, and the logarithmic baseline correction, as well as the
	results without any correction. To be counted, the FFT spectrum must
	pass a signal-to-noise-ratio (SNR) of > 1.0 at the desired periodicity
	pitch frequencies in the range of $\pm$3\,Hz. The SNR in this case is a
	measure how salient the frequency peak of interest is in relation to the
	corresponding frequency value in the baseline.
  \item the induced method: An FFT was applied on each trial, starting 50\,ms
	after stimulus onset, and on its baseline individually. All FFTs were
	averaged. We again proved the outcomes of four types of baseline
	correction mentioned above, as well as the results without any
	correction. To be counted, the FFT spectrum had to pass a SNR > 1.0 at
	the desired periodicity pitch frequencies in the range of $\pm$3\,Hz. 
\end{enumerate}

\section{Results and Conclusions}
\subsection{Low Frequency Analysis}
\begin{figure}
\centering
\includegraphics[width=\linewidth]{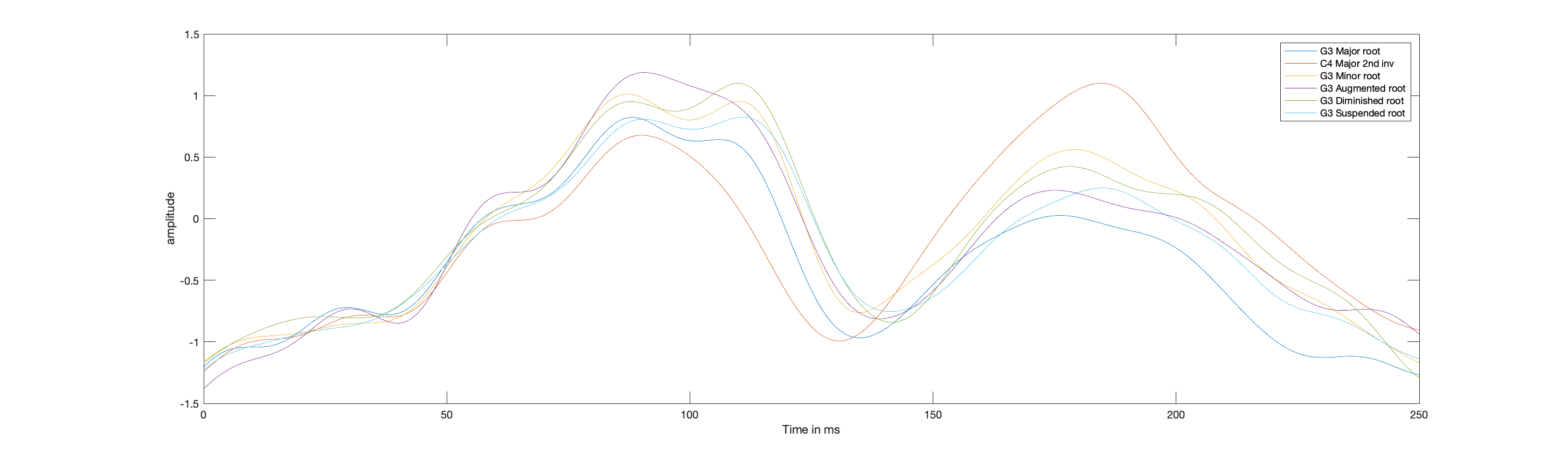}
\caption{This figure shows the EEG response of CZ, filtered in the range
0--30\,Hz, for all six triads. There are clear peaks at N100 and N200. The first
peak is called an "orienting response". It matches with previously experienced
stimuli, whenever a stimulus occurs and has maximum amplitude over Cz
\cite{SS09}.}
  \label{fig:time}
\end{figure}

The results of the low frequency analysis in the time domain show a clear
negative deflection peaking N100 and N200 in all chords, i.e., negative going
event related potentials about 100\,ms and 200\,ms after the onset of the
stimulus (cf. \cref{fig:time}). The N100 peak is a clear evidence for the
presence of an auditory stimulus. Those early waves are observed when an
unexpected stimulus is presented and depend on physical parameters of the
stimulus. The N200 peak shows the mismatch negativity, a component that is
triggered by any discriminable change in a repetitive background of auditory
stimulation and represents an automatic process of the brain occuring while
encoding of a stimulus difference \cite{SS09}. Since all stimuli are permuted in
a random order, the mismatch negativity is found in every response. The C Major
chord shows a significant higher mismatch negativity because its dominant
frequencies are approximately double of all the other triad frequencies.

The analysis of the low frequency area in the frequency domain results in a
hardly noticable difference between all chords. Though the C major chord in
second inversion shows a slightly higher amplitude in brain activity around
10\,Hz in the frontal area, being composed of the electrodes Fp1, Fpz, Fp2, Af3,
Afz, and Af4, the diminished chord dominates the same frequency in the
Cz-electrode. Considering the amplitude average of all triads in the range
9--11\,Hz there is a higher average in the frontal area than in the Cz-electrode
(cf. \cref{tab:amplitude}). Among all examined frequencies and relative
periodicities the C major in the second inversion uses the highest audited
frequencies (cf. \cref{tab:stim}) but the lowest relative periodicity of 3.

\begin{table}[!h]
\centering
\begin{tabular}{lcc}
stimulus interval & $\varnothing$ amplitude Cz & $\varnothing$ amplitude Frontal area
\\\hline
G major root			& 0.062 & 0.0793\\
C major 2nd			& 0.067 & 0.0720\\
G minor root			& 0.0697 & 0.0762\\
G augmented root		& 0.0712 & 0.0738\\
G diminished root		& 0.0707 & 0.0721\\
G suspended root		& 0.0706 & 0.0737\\
\end{tabular}
\caption{Stimuli and the amplitude average in the range 9--11\,Hz.}
\label{tab:amplitude}
\end{table}

\begin{figure}
  \centering
  \begin{subfigure}[b]{0.49\linewidth}
    \includegraphics[width=\linewidth]{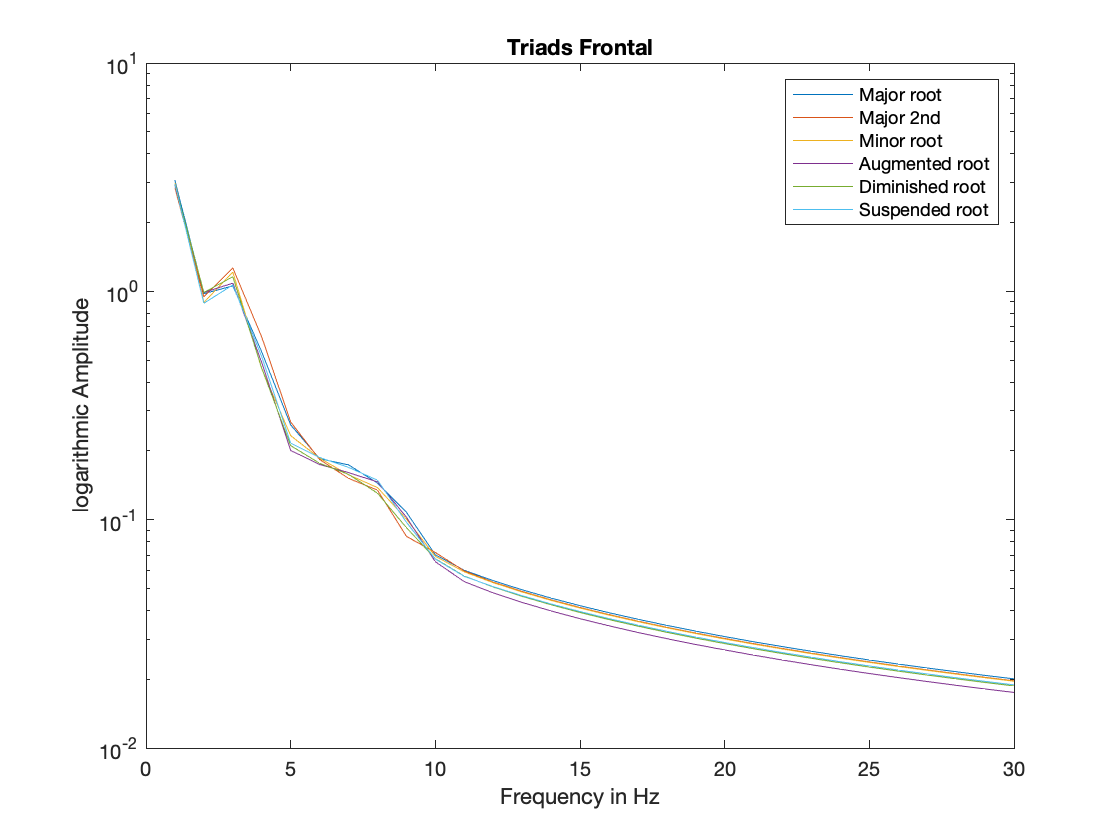}
    \caption{Frontal area}
  \end{subfigure}
  \begin{subfigure}[b]{0.49\linewidth}
    \includegraphics[width=\linewidth]{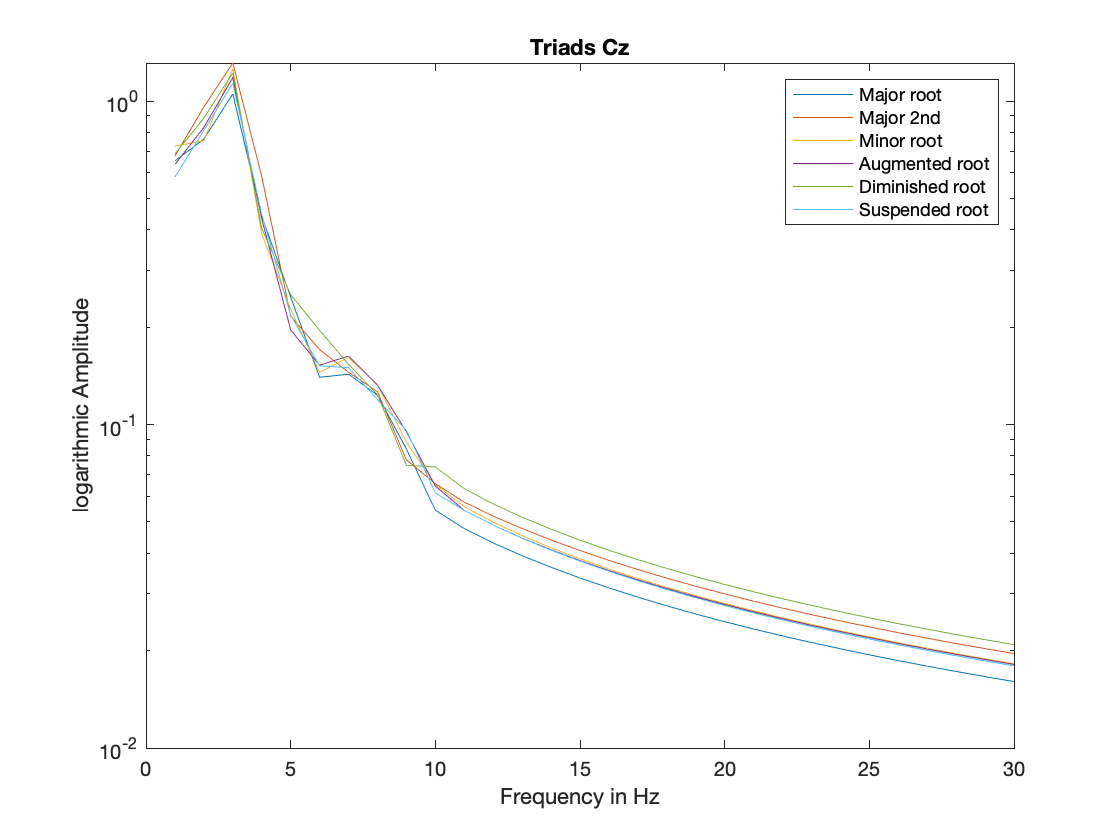}
    \caption{Cz-Area}
  \end{subfigure}
  \caption{This figure shows the frequency spectra in the range 0--30\,Hz of the used triads in a logarithmic scale. The amplitudes around 1--2\,Hz refer to the heartbeat, the values in the area around 5\,Hz consider eye blinks. Heartbeat and eye blinks dominate all other frequencies, the difference in amplitude is insignificant among all chords.}

  \begin{subfigure}[b]{0.49\linewidth}
    \includegraphics[width=\linewidth]{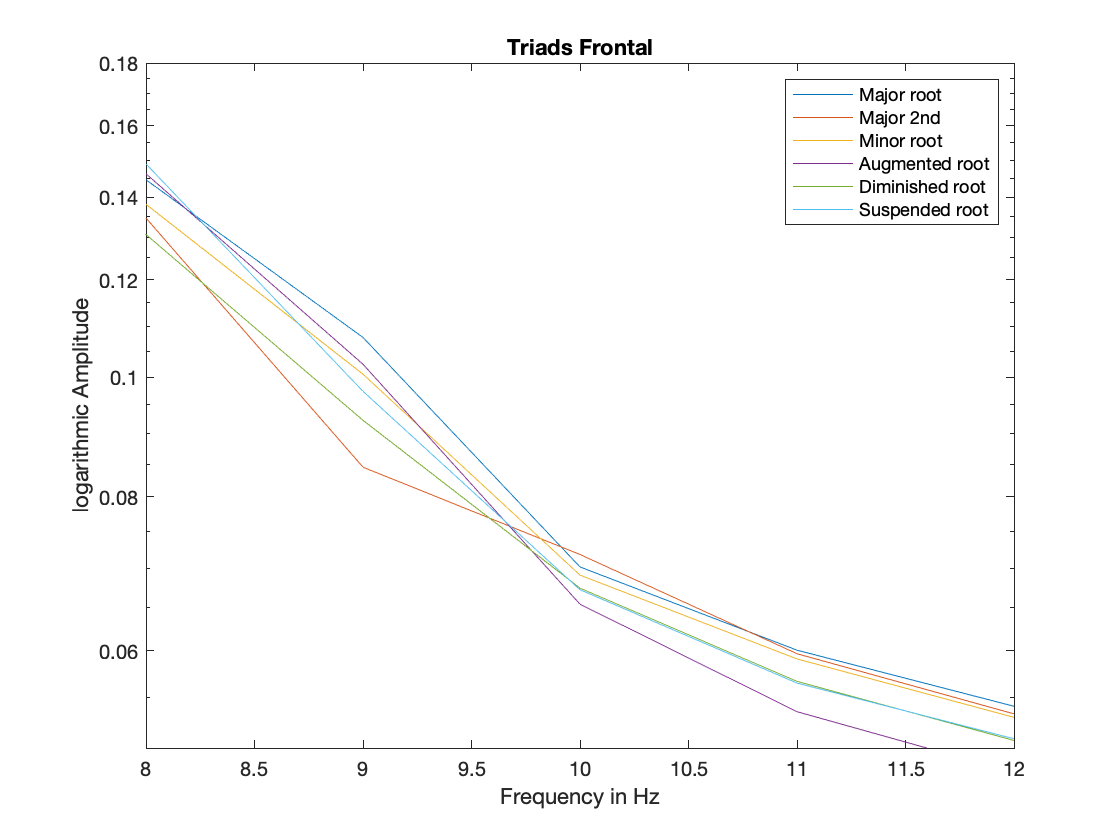}
    \caption{Frontal area}
  \end{subfigure}
  \begin{subfigure}[b]{0.49\linewidth}
    \includegraphics[width=\linewidth]{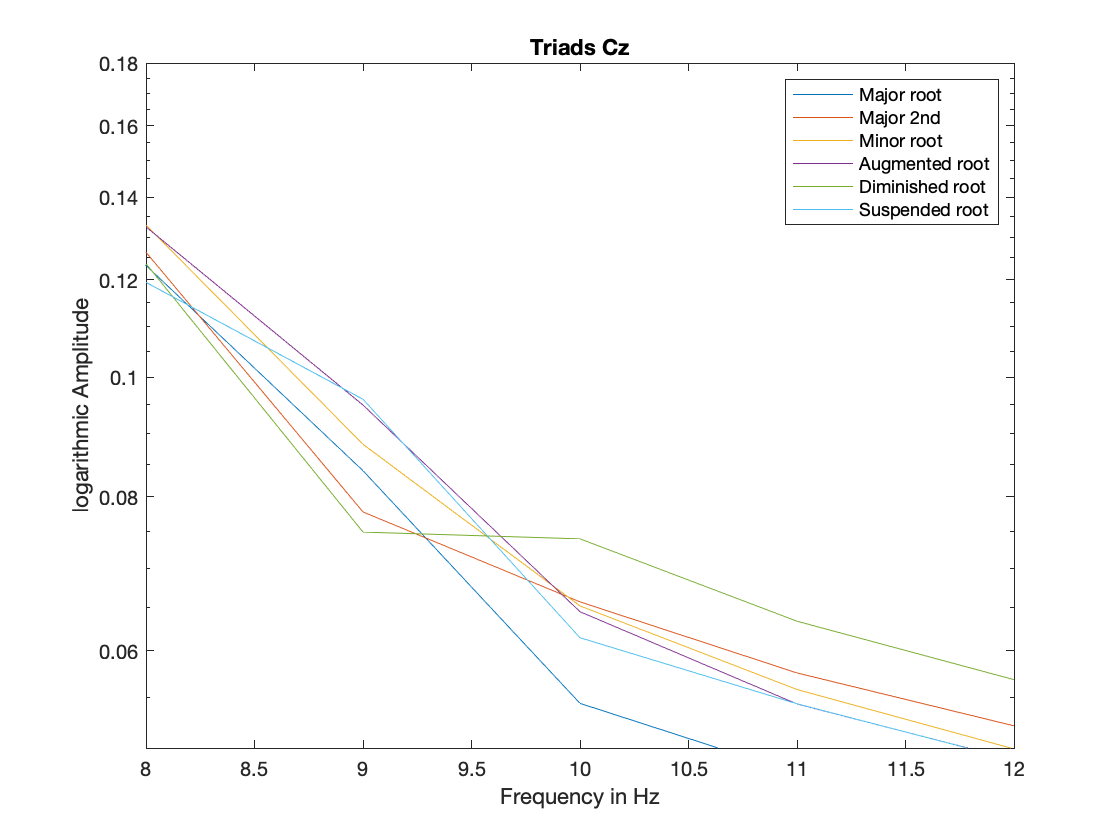}
    \caption{Cz-Area}
  \end{subfigure}
  \caption{This figure shows a more detailed frequency spectra in the range 8--12\,Hz of the used triads. The amplitudes in this area consider brain activity. }
  \label{fig:low}
\end{figure}
  
\subsection{High Frequency Analysis}
Unfortunately, in the auditory brainstem responses we rarely found the
sought-for periodicity pitch frequencies and the stimulus frequencies did not
appear in the frequency spectra at all. There was no difference in the method or
baseline corrections we used. 
For meaningful and correct results the hardware of the used system is quite
important. EEG systems are not jitter-free and it is impossible to find the
correct delay for each trial, because jitter-effects are not of equal durations
and vary from trial to trial. Those problems can be overcome with a parallel but
separate recording of the sound with which the starting point of the auditory
stimulus can be set offline afterwards.

\cite[p.~11]{SK10} recommends that the deviation $\tau$ caused by jitter should
not exceed 0.1\,ms for a properly functioning system. We can derive this order
of magnitude for the jitter as follows: According to the Nyquist-Shannon sampling
theorem, a sampling rate of $f=5$\,kHz allows to detect frequencies up to
$f/2=2.5$\,kHz in principle. Because of the bad SNR ratio in the EEG
experiments, however, the average of many trials has to be considered in order
to reduce noise (cf. \cref{analysis}). If we want to detect the peaks of
cosine components in the averaged signal and assume that the jitter is uniformly distributed around
zero by $\pm\tau$, then we obtain $x = \int_{-\tau}^{+\tau} \cos(\omega t) /
(2\tau)\,dt = \sin(\omega \tau) / (\omega \tau)$ with $\omega = \frac{2\pi}{T}$
and $1/T = f/2$ as expected value for the peaks -- instead of $\cos(0)=1$. We
find the first zero of $x$ for $\omega \tau = \pi$. It should be $x > 0$ and
thus $\tau \ll 1/f = 0.2$\,ms.

Another main issue is the presentation of the sound. One reason might be that
providing the sound through loudspeakers cause a significant attenuation of the
sound intensity. Although loudspeakers should serve its purpose, it would be
good to use magnetic shielded in-ear headphones to also avoid delays or even
possible hearing difficulties of the participants with increasing age. Since the
auditory brainstem shows low-pass characteristics \cite{SK10}, it is
necessary to choose stimuli with low frequencies 80–300\,Hz and instruments with
frequencies in this target range, otherwise the relatively low amplitudes of
those frequencies will be superimposed by noise and will not occur in the spectrum
of the EEG response. The synthesized classic piano sound was possibly not be the
best choice because of its high harmonics. Because the highest frequency of all
stimuli is well below 2.5\,kHz, the sampling rate of 5\,kHz should not be a problem.

Ongoing work includes new experiments with more suitable stimuli and, most
important, the jitter-related problems are fixed by preparing the stimuli in a
different way, namely putting them all in one single audio file. Stay tuned!

\bibliographystyle{plainurl}
\bibliography{frieder,stolzen,extra}

\newpage\appendix
\section{Addendum}

\subsection{Repetition of the EEG Experiments}

The EEG experiments on harmony perception in the human brain reported in \cite{HH+20} (see main part above)
focus on the detection of periodicity pitch in complex harmonies
using EEG data. However, a major challenge identified in the EEG experiments is
jitter in the EEG system, which can result in small timing inaccuracies when
synchronizing the stimulus with the EEG recording. This variability affects the
averaged signal and can reduce the accuracy of detecting peaks in the frequency
spectrum. The study highlights that even small errors in timing can lead to
signal cancellation and introduce noise, making it difficult to interpret the
results.

To mitigate this, one single coherent recording of the stimuli can be used to
offset the jitter problem. With this setting, the EEG experiments on harmony
perception have been repeated with seven stimuli (cf.~\cref{stimuli}) consisting
of 300\,ms synthesized tuba triads with 100\,ms intervals, presented while
participants watched a muted nature documentary. EEG data was collected using 32
active electrodes at a sampling rate of 10\,kHz. The experiment had three blocks
with each block repeating the stimuli 500 times, totaling 15,000 repetitions per
stimulus. Fifteen healthy adult listeners (13 females, 2 males; mean age 21.6
years) without known hearing difficulties participated in this research. From
all participants an informed consent was obtained. The Ethical Review Committee
Psychology and Neuroscience of Maastricht University also approved the new study.

\begin{table}
\centering\small
\begin{tabular}{|l|c|c|c|r@{~}l|}
\hline
& \textbf{Tone} & \textbf{Frequency} & \multicolumn{3}{|c|}{\textbf{Amplitude}}\\
\textbf{Stimulus Chord} & \textbf{Pitch} & \textbf{[Hz]} & \textbf{Stimulus} & \multicolumn{2}{|c|}{\textbf{Response}}\\
\hline
G major (root) & PP & 49 (98) & 0.0017 (0.0011) & --&(0.0046)\\
 & G3 & 197 (393) & 0.0523 (0.0410) & 0.0088&(0.0095)\\
 & B3 & 248 (494) & 0.0619 (0.0435) & 0.0148&(0.0077)\\
 & D4 & 295 (588) & 0.0692 (0.0233) & 0.0064&(0.0022)\\
\hline
C major (2nd inv.) & PP & 66 (131) & 0.0016 (0.0011) & 0.0030&(0.0015*)\\
 & G3 & 197 (393) & 0.0572 (0.0447) & 0.0103&(0.006*)\\
 & C4 & 262 (524) & 0.0782 (0.0445) & 0.0108&(0.005)\\
 & E4 & 331 (660) & 0.0621 (0.0421) & 0.0175&(0.004)\\
\hline
G minor (root) & PP & 20 (39) & 0.0013 (0.0017) & --&(0.0029)\\
 & G3 & 197 (393) & 0.0560 (0.0437) & 0.0113&(0.0103)\\
 & B$\flat$3 & 234 (467) & 0.0604 (0.0405) & 0.0131&(0.0066)\\
 & D4 & 295 (588) & 0.0741 (0.0254) & 0.0104&(0.0032)\\
\hline
G diminished (root) & PP & 39 (78) & 0.0017 (0.0010) & --&(--)\\
 & G3 & 197 (393) & 0.0600 (0.0465) & 0.0123&(0.0093)\\
 & B$\flat$3 & 234 (467) & 0.0647 (0.0434) & 0.0133&(0.0068)\\
 & D$\flat$4 & 278 (555) & 0.0585 (0.0352) & 0.0106&(0.0060)\\
\hline
G suspended (root) & PP & 33 (66) & 0.0015 (0.0017) & 0.004&(0.0043)\\
 & G3 & 197 (393) & 0.0559 (0.0443) & 0.0115&(0.0112)\\
 & C4 & 262 (524) & 0.0772 (0.0439) & 0.0133&(0.0076)\\
 & D4 & 295 (588) & 0.0736 (0.0255) & 0.0077&(0.0031)\\
\hline
G suspended (1st inv.) & PP & 25 (49) & 0.0010 (0.0016) & --&(0.0053)\\
 & G3 & 197 (393) & 0.0458 (0.0336) & 0.0101&(0.0078)\\
 & A3 & 221 (440) & 0.0810 (0.0402) & 0.0172&(0.0075)\\
 & D4 & 295 (588) & 0.0594 (0.0199) & 0.0076&(0.0033)\\
\hline
G suspended (2nd inv.) & PP & 13 (26) & 0.0014 (0.0015) & --&(0.005*)\\
 & G3 & 197 (393) & 0.0557 (0.0435) & 0.0089&(0.0039)\\
 & C4 & 262 (524) & 0.0765 (0.0430) & 0.0072&(0.0063)\\
 & F4 & 350 (699) & 0.0580 (0.0442) & 0.0211&(0.0049)\\
\hline
\end{tabular}
\par
\hspace*{\fill}{* no global peak}

\caption{This table shows the seven stimuli and the corresponding frequencies.
Each stimulus consists of the triad frequencies as well as their second partial
frequencies (in brackets). The suspended chord in root position used in this
experiment contains a perfect fourth. PP means periodicity pitch and, as can be
seen from the low values of the column with stimulus amplitudes, is not included
in the stimulus frequencies.}
\label{stimuli}
\end{table}

\subsection{Main Research Question and Related Works}

The main research question addressed with the new EEG experiments is again
whether the brain adds periodicity pitches, which are not physically present in
the auditory stimuli, through the Frequency Following Response (FFR). For the
sake of completeness, let us recall that the periodicity pitch of a complex
chord is related to its missing fundamental and can be derived from its relative
periodicity $h$ that is the approximated ratio of the period length of the whole
chord to the period length $t$ of its lowest tone component \cite{Sto15}. The
periodicity pitch can then be calculated as $f = 1/(h \cdot t)$.
As \cite{Sto15} shows, the perceived consonance of a harmony decreases as the
relative periodicity increases.

As known, several researchers investigate the auditory brainstem response to
musical harmonies. There are EEG experiments with several musical intervals
(i.e., dyads) \cite{LS+09,LS+15} and also with triads \cite{BK11} as stimuli.
The additionally occurring frequencies reported in the studies coincide well
with the periodicity pitch frequencies of the respective musical harmonies.
Therefore \cite{Sto17b} aims to find out more precisely the relevant factors
that lead to the occurrence of the periodicity pitch in the response spectrum of
a signal and investigates the transformation of the input signal into pulse
trains (spikes) whose maximal amplitude is limited by a fixed uniform value.

\subsection{Results and Conclusion}

\begin{figure}
\begin{subfigure}[b]{0.42\linewidth}
	\includegraphics[width=\linewidth]{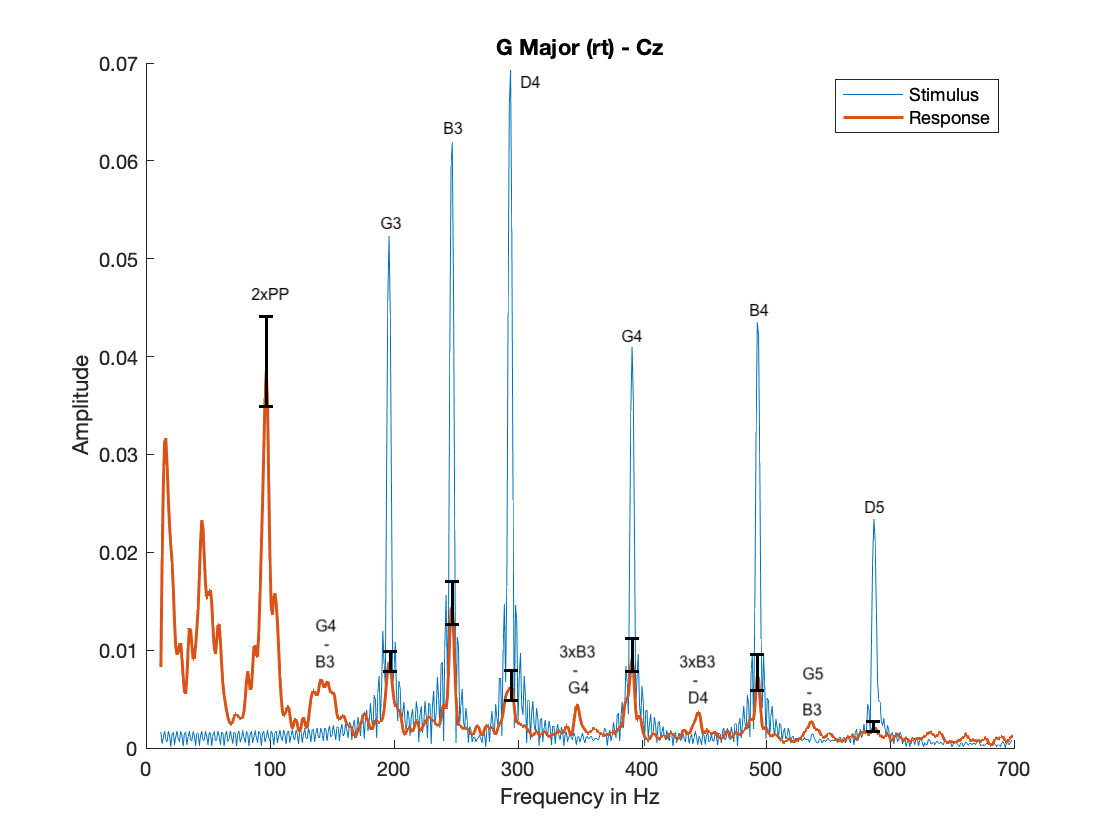}
	\caption{G major, root position.}
\end{subfigure}
\begin{subfigure}[b]{0.42\linewidth}
	\includegraphics[width=\linewidth]{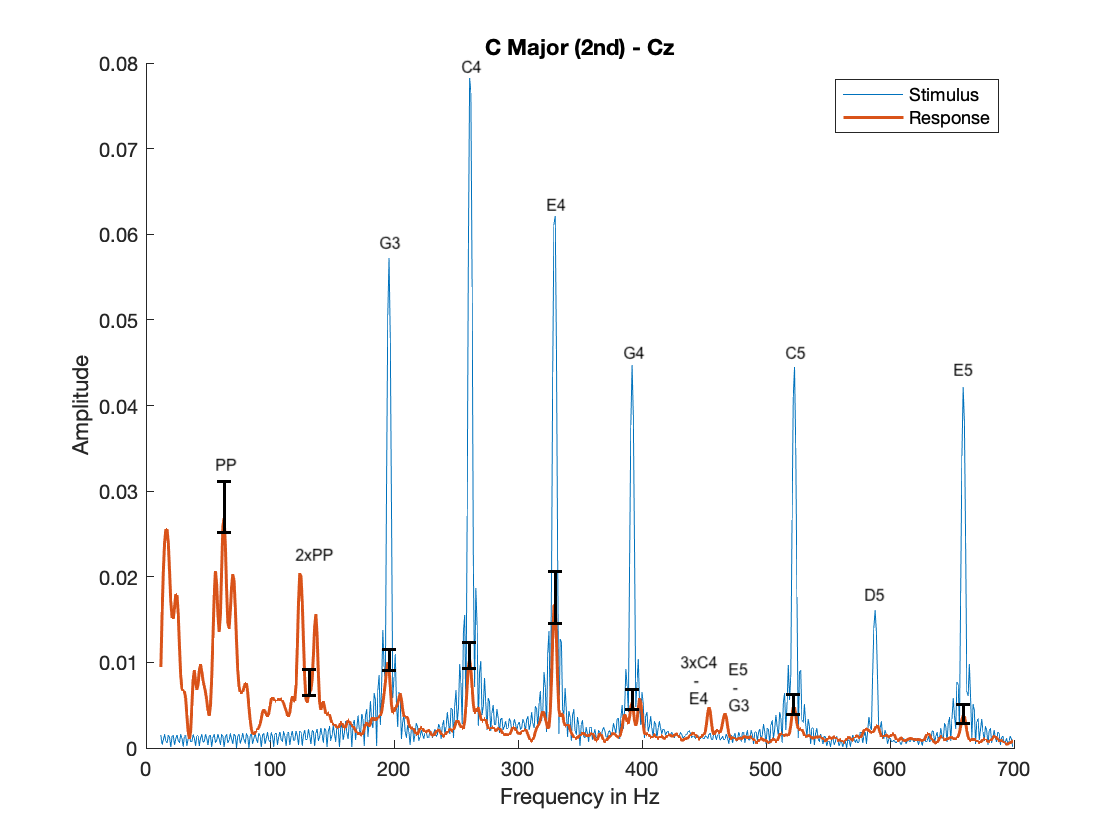}
	\caption{C major, second inversion.}
\end{subfigure}
\begin{subfigure}[b]{0.42\linewidth}
	\includegraphics[width=\linewidth]{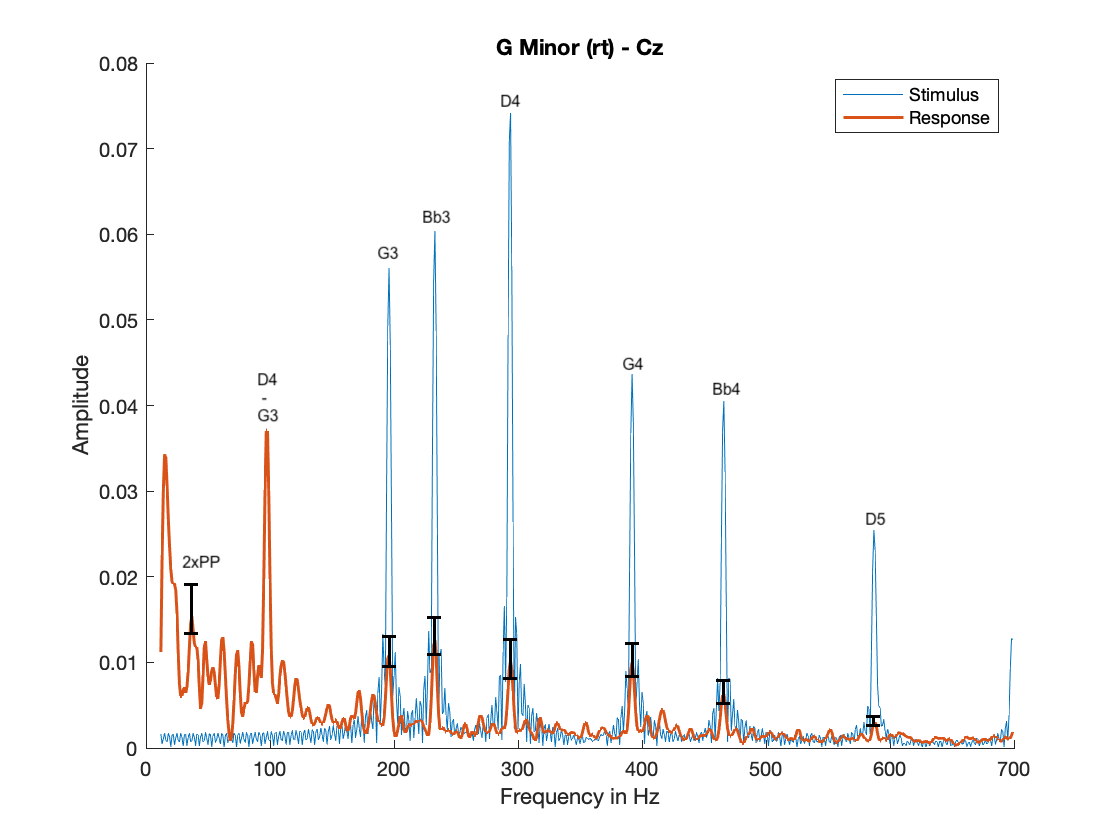}
	\caption{G minor, root position.}
\end{subfigure}
\begin{subfigure}[b]{0.42\linewidth}
	\includegraphics[width=\linewidth]{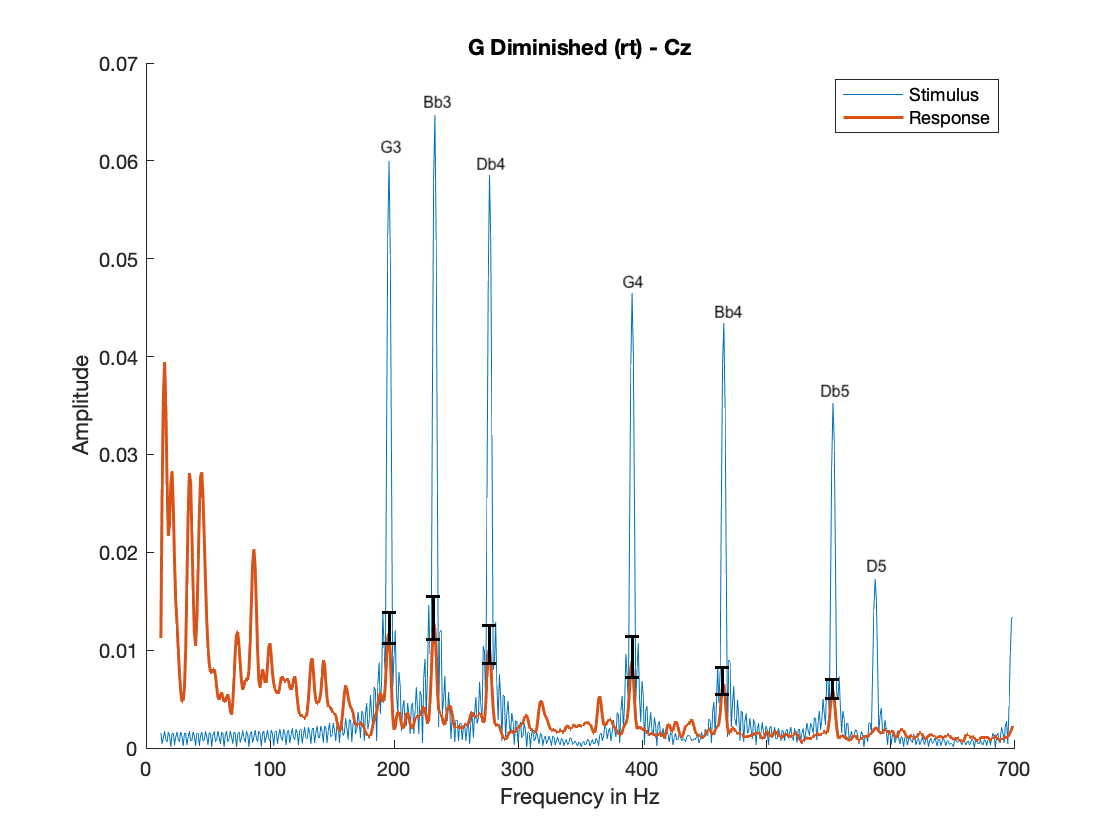}
	\caption{G diminished, root position.}
\end{subfigure}
\begin{subfigure}[b]{0.42\linewidth}
	\includegraphics[width=\linewidth]{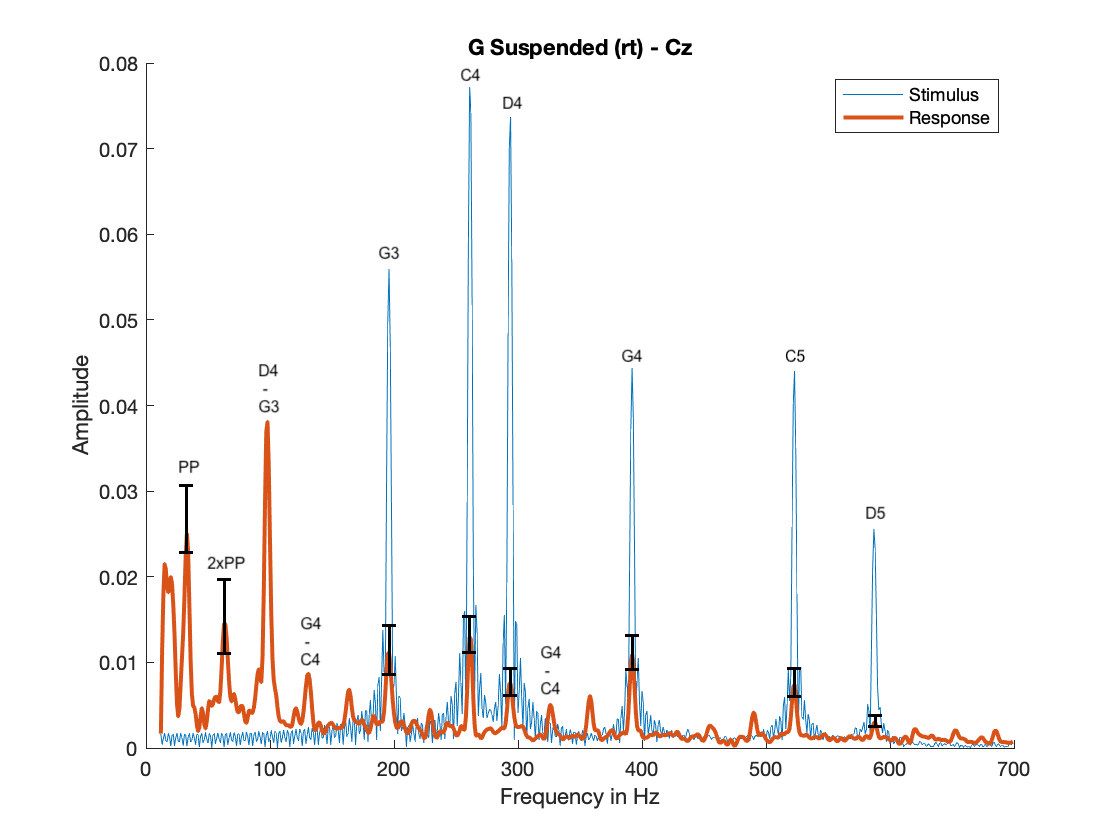}
	\caption{G suspended, root position.}
\end{subfigure}
\begin{subfigure}[b]{0.42\linewidth}
	\includegraphics[width=\linewidth]{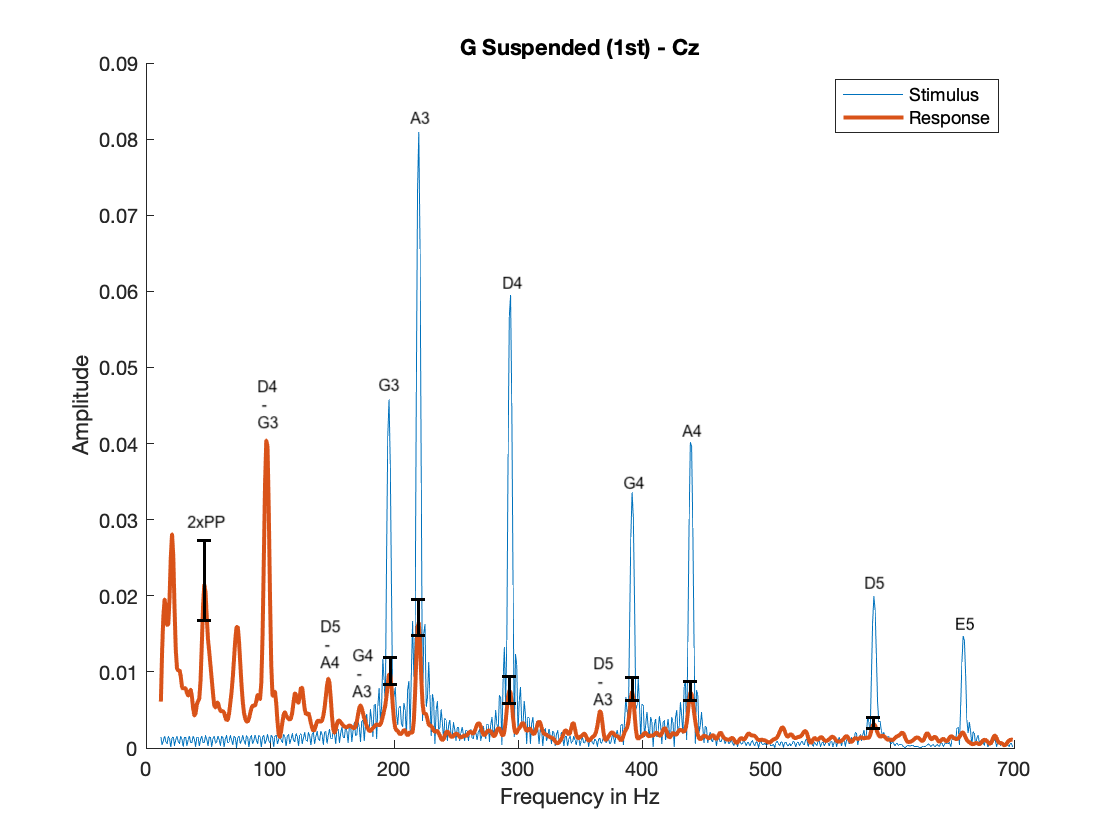}
	\caption{G suspended, first inversion.}
\end{subfigure}
\begin{subfigure}[b]{0.42\linewidth}
	\includegraphics[width=\linewidth]{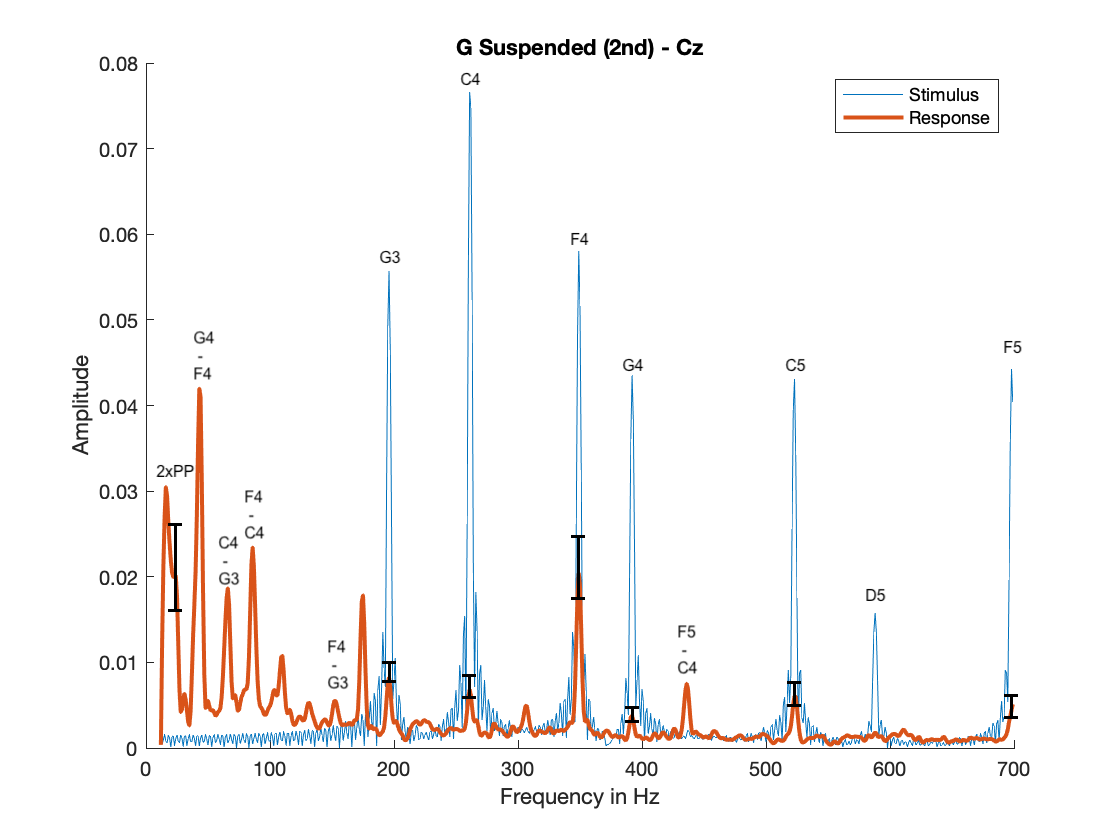}
	\caption{G suspended, second inversion.}
\end{subfigure}
\caption{FFRs for the seven stimuli.}\label{ffrs}
\end{figure}

The results of our new EEG experiment confirm that the periodicity pitch (PP) or
its double can be detected in the FFR at the Cz electrode for considered triads
in most cases, except for the diminished chord in root position and the second
inversion of the suspended chord. The frequencies of the stimuli and some of its
partials are also present in the FFR (cf.~\cref{ffrs}). For instance, the
suspended chord in root position shows peaks in the FFR for the periodicity
pitch (33\,Hz) and its double as well as the simulus frequencies (i.e., of G3,
C4, D4) including difference tones (e.g., G4-C4 and its double).

Nevertheless, for some stimuli, the periodicity pitch is masked by nearby
frequencies, or it fell outside the detectable range due to filtering. Small
periodicity pitch frequencies (e.g., 20\,Hz for the minor chord in root
position) should be avoided anyway, because they are close to low frequency
activities like muscle movements. The reported results must therefore be
regarded as provisional, also due to the small number of participants. Thus in
further future work, the experiments should be repeated once again with more
participants and higher frequencies in the stimuli.

\end{document}